\documentstyle[osa,aps,prb,epsfig]{revtex}
\begin{document}
\twocolumn[\hsize\textwidth\columnwidth\hsize\csname
@twocolumnfalse\endcsname
\title{
Small Josephson current and product deduced by means of
measurement for an inhomogeneous superconductor: Extension of the
Ambegaokar - Baratoff theory}
\author{Hyun-Tak Kim (htkim@etri.re.kr, ~kimht45@hotmail.com)}
\address{Telecom. Basic Research Lab., ETRI, Daejeon 305-350, Korea}
\maketitle{}
\newpage
\begin{abstract}
For an inhomogeneous high-$T_c$ superconductor, band-filling
dependence of Josephson current and Josephson product is deduced
at $T$=0 K by means of measurement; this is an extension of
Ambegaokar - Baratoff Josephson current and product. The observed
Josephson current, $J_{obs}$, is given by $J_{obs}={\rho}J_i$,
where 0$<\rho\le$1 is band filling (or local density). When
$\rho$=1, $J_{obs}$ = $J_i$  is the intrinsic supercurrent
occurring by Cooper pair. When 0$<\rho<$1, $J_{obs}$ is an average
of $J_i$ over the measurement region and is the effect of
measurement. The observed Ambegaokar - Baratoff Josephson product,
based on the $s$-wave theory, is given by $J_{obs}R_n
=\frac{\pi}{2}\rho\triangle_i$, where $\triangle_i$ is the
intrinsic superconducting gap and small, which results in small
Josephson products observed by the experiments. The intrinsic gap,
4$<{\triangle_i}<$10 meV, is analyzed from the Josephson-product
data of Bi$_2$Sr$_2$CaCu$_2$O$_{8+x}$. In addition, the
triple-${\pi}$-junction experiments, observing the half-flux
quantum (${\Phi_0}/2$) as evidence of $d$-wave symmetry, are
discussed by using means of measurement. Comments on a flux trap
of the magnetic modulation experiment using Pb-YBCO dc SQUIDs are
also given. Author's thought is given on the angle dependence of
$J_c$ measured in whisker.
\\ PACS numbers: 74.20.Fg, 74.50.+r \\ \\
\end{abstract}
]
From the discovery of a high-$T_c$  superconductor until recently,
pairing symmetry for the mechanism of high-$T_c$ superconductivity
has been controversial because intrinsic physical information of
superconductors is  not obtained for intrinsically inhomogeneous
superconductors$^{1-5}$ with a metal phase and an insulator phase
with the $d_{x^2-y^2}$-wave symmetry.$^6$ The intrinsic
inhomogeneity in which a homogeneous metal region is about 14$\AA$
was revealed by scanning tunnelling microscopy.$^1$ Moreover, as
another evidence of the inhomogeneity, coexistence of the
superconducting gap and the pseudo-gap was observed.$^{7,8}$ The
inhomogeneity is due to the metal-insulator instability.$^5$ In
inhomogeneous superconductors, the fact that the energy gap
decreases with an increasing local density was also revealed by
experiments$^{1-4}$ and a theoretical consideration$^6$. Recently,
for an inhomogeneous superconductor, an analysis method for the
intrinsic density of states$^{9,10}$ and the intrinsic
superconducting gap$^6$ was developed by means of measurement. The
method disclosed the identity of gap anisotropy and revealed that
pairing symmetry of a high-$T_c$ superconductor is an
$s$-wave.$^6$

However, there  are  still two  unsolved problems to be clarified.
One is that the Josephson product decreases  with an increasing
superconducting gap.$^{11}$ This also creates a problem in
determining the extent of the $s$-wave component in $s+d$ mixed
pair states. Many investigations indicated that the Josephson
product in c-axis Josephson pair (or intrinsic Josephson)
tunneling experiments is much smaller than the all-$s$-wave
Ambegaokar-Baratoff limit.$^{12-20}$ On the basis of this
experimental result, it has been interpreted that the ratio of the
$s$-wave component to the full $d$-wave one in the $s+d$ mixed
states is very small.

The other problem is a doping dependence of the Josephson
supercurrent for inhomogeneous superconductors. Tsuei
$et~al.^{21-23}$ and Sugimoto $et~al.^{24}$ observed the half flux
quantum as evidence of the $d$-wave symmetry for a thin film
deposited on a tricrystal substrate. However, this result was
analyzed by assuming that the films are homogeneous
superconductors. Related to pairing symmetry, the problems are
very important for the mechanism of high-$T_c$ superconductivity.

In this paper, we deduced band-filling (or doping) dependence of
the Josephson supercurrent and the Josephson product at $T$=0 K,
by using the means of measurement suggested in a previous
paper$^{6,9}$; this is an extension of Ambegaokar - Baratoff
Josephson current and product. In addition, the
triple-${\pi}$-junction experiments for observing the half-flux
quantum are discussed by using a Josephson current developed here.
Comments on flux trap of the magnetic modulation experiment using
Pb-YBCO dc SQUIDs are also given.

Fractional charge has been demonstrated in previous
papers.$^{6,9}$ These will be reviewed briefly. When an
inhomogeneous superconductor with two phases of a metal region and
an insulating region$^6$ is measured, carriers in the metal region
are averaged over lattices (or atoms) in the entire measurement
region, including the two-phase regions. The metal region has the
electronic structure of $one~electron~per~atom$. Then the
inhomogeneous superconductor is changed into a homogeneous one
with the electronic structure of $one~effective~charge~per~atom$.

The observed effective charge becomes $e'={\rho}e$, where
0$<{\rho}=n/l{\le}$1 is band filling, $n$ is the number of
carriers in the metal region, and $l$ is the number of total
lattices in the measurement region. The fractional effective
charge is justified only when the inhomogeneous system is
measured. Otherwise, it becomes true charge in the metal region
(see reference 9). By using the fractional charge, in   an
inhomogeneous superconductor, the observed energy gap,
$\triangle$, was given by

\begin{eqnarray}
{\triangle}={\triangle}_i/{\rho} ,
\end{eqnarray}
where $\triangle_i$ is the intrinsic  superconducting gap
determined by the minimum bias voltage. The 0$<{\rho}{\le}$1 is
band filling (local density or inhomogeneous factor), which
indicates the extent of the metal region.$^6$ The validity of Eq.
(1) was given by many tunneling experiments.$^6$

Ambegaokar and Baratoff$^{25}$  derived the Josephson  tunnelling
supercurrent on the basis of BCS theory for  an $s$-wave
homogeneous superconductor. The suppercurrent at $T$=0 K was given
by

\begin{eqnarray}
J = \frac{\pi}{2}R_n^{-1}{\triangle},
\end{eqnarray}

where $R_n = (2{\pi}h/e^2T)$, $T$ is the tunneling matrix, and
$\triangle$ is a superconducting energy gap.

In an inhomogeneous superconductor, the averaged metallic system
has the electronic structure of one effective charge per atom,
which is mathematically equivalent to the electronic structure of
the metal used in the BCS theory. The metal for k-space used in
the BCS theory has the electronic structure of one electron per
atom. The Josephson current and product derived in the Ambegaokar
and Baratoff theory can be used without formula's change even in
the inhomogeneous superconductor by replacing true charge by the
effective charge because the averaged effective charge is
invariant under transformation. In addition, a similar calculation
has been given when the Brinkman-Rice picture was extended (see
reference 9). Thus, the observed supercurrent, $J_{obs}$, is given
by substituting $e$ in $R_n$ and $\triangle$ with $e'={\rho}e$ and
$\triangle$=${\triangle_i}/{\rho}$ by
\begin{eqnarray}
J_{obs} {\equiv}(\frac{e^2T}{4h}){\rho}{\triangle_i}={\rho}J_i,
\end{eqnarray}
where $J_i$ is the intrinsic supercurrent. The observed Josephson
product is also given by using Eq. (3) by

\begin{eqnarray}
J_{obs}R_n
 {\equiv} \frac{\pi}{2}{\rho}^2{\triangle_{obs}}{\equiv}\frac{\pi}{2}{\rho}{\triangle_i}={\rho}J_iR_n,
\end{eqnarray}

where $\triangle_i$ is constant.

When ${\rho}$ =1, Eq. (3) and Eq. (4) are the intrinsic
supercurrent and the intrinsic product (or Ambegaokar and Baratoff
product) occurring by true charges, respectively. When
0$<{\rho}<$1, the equations correspond to the averages of the
intrinsic supercurrent and the intrinsic product over the
measurement region and are the effect of measurement. Eq. (3) and
Eq. (4) are the extended Ambegaokar - Baratoff Josephson current
and product, respectively.

The Josephson current and the Josephson product decrease with
decreasing ${\rho}$. This explains the fact that the observed
Josephson product decreases with an increasing energy gap.$^{11}$
The ${\rho}$ dependence in Eq. (4) comes from Eq. (3), which
agrees with a result observed by the intrinsic Josephson
junction.$^{17}$ Note that the magnitude of the intrinsic product
is basically very small because ${\triangle_i}$ is small.
Considering that the anisotropy of the number of carriers in the
c-axis and ab-plane is large
(${\rho_{c-axis}}<<{\rho_{ab-plane}}$), the product observed in
the c-axis is naturally much less than that in the ab-plane. Thus,
the observed small Josephson products$^{12-20}$ can be explained
by Eq. (4). Additionally, for a Josephson junction by two
superconductors with different energy gaps, the Josephson product
derived by Anderson$^{26}$ is in the context of the above
analysis.

We analyze the intrinsic gap of Bi$_2$Sr$_2$CaCu$_2$O$_{8+x}$ from
experimental data, using Eq. (4). Irie $et~al.^{17}$ suggested
that $I_cR_n\approx$13.3 meV observed by the intrinsic Josephson
junction is $\frac{1}{3}$ of the Ambegaokar-Baratoff product,
$I_cR_n\approx$ 40 meV, using $\triangle_{obs}\approx$25 meV and
Eq. (4) with $\rho$=1. The true Ambegaokar-Baratoff product is
much less than 40 meV, when both $\rho\ne$1 and
$\triangle_{obs}\ne\triangle_i$ are considered. The intrinsic gap,
$\triangle_i\approx$8.5 meV, is obtained from the observed
product, $I_cR_n\approx$13.3 meV by using Eq. (4) with $\rho$=1.
The true intrinsic gap is slightly larger than 8.5 meV because
$\rho<$1 slightly. Miyakawa $et~al.^{14}$ measured a
Josephson-product, $J_{obs}R_n{\approx}$7 $meV$ for an over-doped
crystal, which can be regarded as $\rho\approx$1 without a
pseudogap$^{27}$. The intrinsic gap, $\triangle_i$=4$\sim$5 $meV$,
is obtained by Eq. (4). Mourachkine$^{11}$ observed the average
Josephson product, $I_cR_n\approx$ 10 meV, at the minimum energy
gap. The intrinsic gap, $\triangle_i\approx$~6.4 meV, is obtained
by Eq. (4) with $\rho$=1. The true intrinsic gap is slightly
greater than 6.4 meV because $\rho<1$ slightly. Thus, we conclude
that the analyzed intrinsic gap is in 4 $<{\triangle_i}<$10 meV.

We apply Eq. (3) to the Josephson-${\pi}$ junction theory.$^{28}$
The observed circulating current in an inhomogeneous
superconducting ring is given when $I_{ij}L>{{\Phi}_0}$ without
any external field$^{21}$ by

\begin{eqnarray}
I_{obs} {\equiv}{\rho}I_i{\approx}{\rho}{\frac{\it\Phi_0}{2L}},
\end{eqnarray}

where ${\Phi}_0$ (=$h/2e$) is the flux quantum and $L$ is a
self-inductance of a superconducting ring. When 0$<{\rho}<$1, the
observed supercurrent, $I_{obs}$, is an average of the intrinsic
supercurrent over the measurement region.

Tsuei $et~al.^{21-23}$ and Sugimoto $et~al.^{24,29}$ analyzed the
half-flux quantum from a flux observed by a scanning
superconducting-quantum-interference-device (SQUID) loop for
frustrated (or $\pi$ loop) YBa$_2$Cu$_3$O$_{7-\delta}$ (YBCO)
films deposited on a tricrystal substrate, and emphasized the
observation of the half-flux quantum as evidence of the $d$-wave
symmetry. However, they did not consider ${\rho}$ in Eq. (5) in
their analysis of the experimental data; they assumed $\rho$=1.
The $\rho$ effect differs from the disorder effect applied by
Tsuei $et~al.^{21}$. We regard the YBCO films as inhomogeneous
superconductors of 0$<{\rho}<$1. When $L$, evaluated by Tsuei
$et~al.$, is used, an intrinsic flux to be evaluated,
${\Phi}_i$/2, should be larger than ${\Phi}_0$/2;
${\Phi}_i={\frac{1}{\rho}}{\Phi}_0>{\Phi}_0$.$^{30}$ When
${\Phi}_i$=${\Phi}_0$ and ${\rho\ne}1$ in Eq. (5), a flux value to
be observed should be less than ${\Phi}_0$/2. The half-flux
quantum can be observed only when $\rho$=1. Note that $L$ is
independent of $\rho$ which comes from the supercurrent. In
contrast, their analyzed flux values have a smaller error,
(${\frac{1}{\rho}}-1)({\Phi}_0/2)$, than the expected intrinsic
flux value. To estimate $\rho$ near the triple junction, it is
thought that the characteristic of the superconducting ring is not
good due to the presence of surface stresses on the substrate at a
triple junction which was revealed as either concentrated or
deconcentrated$^{31,32}$. As evidence, the frustrated YBCO film
around a triple junction does not show the Meissner effect
(expulsion of flux) in Fig. 17 (c) of reference 22. Thus, the
inhomogeneity effect cannot be ignored because $\rho$ in the ring
is smaller than that at a region away from junctions. The true
observed flux is about $\frac {1}{3}\Phi_0$ not $\frac
{1}{2}\Phi_0$, as shown in Fig. 2 (a) of reference 23, which
indicates $\rho\ne$1. Nielsen $et~al.^{33}$ also observed the flux
value less than 0.3$\Phi_0$ for the paramagnetic Meissner effect
(PME; single-$\pi$ junction) in Nb-Al$_2$O$_3$-Nb
Josephson-junction arrays, which indicates $\rho\ne$1. Finally,
the intrinsic flux to be evaluated can be regarded as $\frac
{1}{2}\Phi_0$. In addition, Tsuei and Kirtley$^{22,34}$ also
measured the half-flux quantum for  ring, disk and blanket films.

However, Kirtley $et~al.^{23}$ and Sugimoto $et~al.^{24}$ observed
the change of flux direction near $T_c$, which is not explained by
the $\pi$-junction theory, and also observed in the PME of
Ba$_{1-x}$K$_x$BiO$_3$(BKBO)$^{35}$, Nb$^{36}$, YBCO$^{37}$
superconductors. Note that the change of the flux direction
indicates that the magnetic property of a superconductor changes
from paramagnetic (or ferromagnetic) to diamagnetic. Koblischka
$et~al.^{37}$ suggested that the PME in an artificial granular
YBCO film is attributed to field trapping on the basis of the
observation of the upturn near $T_c$. Nielsen $et~al.^{33}$ and
Leo $et~al.^{38}$ suggested that the PME can arise from magnetic
screening in multiply connected superconductors. These are reasons
why we do not think that the observed half flux occurs by the
triple-$\pi$ junctions in the superconducting ring.

We also comment on a flux trap of the magnetic modulation
experiment$^{39}$ using Pb-YBCO dc SQUIDs, on the basis of our
observation of the PME for BKBO$^{35}$. Wollman $et~al.^{39,40}$
mentioned that the trapping of magnetic vortices is observed in
the experiment, but does not affect the corner measurement.
Klemm$^{41}$ commented on the singular demagnetization effects
associated with the corners of very thin samples. We suggest that
what has influence on results at the corner measurement is a
trapped flux but not its quantity, because the PME is observed
although its quantity is very small.$^{35}$ The flux trap is
independent of sample geometry and size and is observed even in
grains of a size of 1${\mu}$m.$^{35}$ For a polycrystal and a
film, a flux is trapped easily, but comes out easily when an
external field disappears. On the contrary, for a high quality
single-crystal, it is difficult to trap a flux.$^{35}$ However,
once a flux is trapped even though the quantity is small, the flux
remains closed in the crystal even when not in an external
field.$^{35}$ This causes the effect of the ${\pi}$ junction, the
half flux due to the ${\pi}$-phase shift. Because Wollman's
experiments measure minimum resistances (or maximum currents) with
varying flux,$^{39}$ the flux can be trapped easily into a sample
even in a very low field. Note that the superconducting state
measuring the maximum currents, as explained in Fig. 1 (b) of
reference 39, is not a perfect condensed state; the
superconductors do not perfectly shield or expel a flux. Thus, a
flux in the YBCO crystal rather than the Pb film can be trapped.
This can also account for the minimum current, a small V shape, or
the maximum resistance, a small peak, near zero flux for single
modulation experiments$^{39}$. This analysis is different from
Wollman $et~al.$'s comments$^{40}$. In addition, the PME and the
half-flux quantum due to the ${\pi}$-phase shift have been
predicted simultaneously in the $\pi$-junction theory.$^{22,28}$
Nevertheless, it has been revealed that the PME occurs by a
trapped flux$^{35-37}$.

In addition, Klemm$^{42}$ shows that a tightbinding $\xi (k)$, to
which the anisotropic extended $s$-wave order-parameter is
applied, fits well into Takano whisker data$^{43}$ in contrast to
Li data$^{44}$. Takano data are the angle dependence of the c-axis
Josephson current suggested as evidence of the $d$-wave symmetry.

This author found that Li's whisker$^{44}$ was annealed in an
enough oxygen atmosphere, while Takano's whisker$^{43}$ in not
enough. Li's whisker can be regarded as an over-doped whisker, but
Takano's whisker as an under-doped one which can have an
insulating phase with a pseudogap. This author wonders whether the
angle dependence of $J(\theta)$ measured by Takano $et~al.$ came
from the superconducting phase or the insulating phase. To prove
this, whisker should be made in an enough oxygen atmosphere. Then
the whisker can show more the intrinsic $J(\theta)$ because an
overdoped crystal has only slight or no insulating phase$^{27}$.
Thus, this author thinks that the Li data without the angle
dependence of $J(\theta)$ indicating $s$-wave symmetry is rather
intrinsic than the Takano data.

In conclusion, Ambegaokar - Baratoff Josephson current and product
were extended. For inhomogeneous high-$T_c$ superconductors,
without adding the $d$-wave theory, Eq. (4) based on the $s$-wave
theory can explain the small Josephson products observed by
experiments. Furthermore, the flux-measurement experiments and the
flux-modulation experiments are not suited to proving the $d$-wave
symmetry because the perfect elimination of the trapped flux is
not possible in the experiments.

\section*{ACKNOWLEDGEMENTS}
The author acknowledges R. A. Klemm for his valuable comments.


\begin{references}
\bibitem[30]{} The flux in the right hand of Eq. (5) is equal to the half-flux quantum observed by experiments;
${\rho}({\Phi}_i/2)=({\Phi}_0/2)_{obs}$, where ${\Phi}_i$ is an
intrinsic flux to be evaluated and $({\Phi}_0/2)_{obs}$ is a value
observed by Tsuei $et~al.$ and Sugimoto $et~al.$. Thus
${\Phi}_i={\frac{1}{\rho}}{\Phi}_0>{\Phi}_0$ due to 0$<\rho<$1.
\end{references}
\end{document}